\documentclass[11pt,a4paper]{article}
\usepackage{jheppub}

\usepackage{amssymb,amsmath}
\usepackage{amsfonts,amsthm}
\usepackage[vcentermath]{youngtab}
\usepackage{tikz}
\usepackage{float}
\usepackage{pgfplots}

\usetikzlibrary{shapes.misc,mindmap,backgrounds,decorations.pathreplacing,decorations.markings}
\usetikzlibrary{decorations.pathreplacing,decorations.pathmorphing,calc}
\usetikzlibrary{patterns}

\makeatletter
\g@addto@macro\@floatboxreset\centering
\makeatother

\title{The Quantum Origins of Non-Topological Vortices}
\author{Carl Turner}
\affiliation{Department of Applied Mathematics and Theoretical Physics, \\
University of Cambridge, \\ 
Cambridge, CB3 0WA, UK}
\emailAdd{C.P.Turner@damtp.cam.ac.uk}
\abstract{We review some unusual facts about the theory of non-relativistic anyons in 2+1 dimensions, and use it as a laboratory to explore how interesting features of non-relativistic field theory correspond to those of many-body quantum mechanics. In particular, we offer an explanation of how Jackiw-Pi vortices arise as the classical limit of certain many-body states in the quantum mechanical theory. Along the way, we make various interesting observations about universal features of the spectrum of anyons subject to different amounts of tuning.}

\def\cO{{\cal O}}

\def\rmd{{\rm d}}

\def\tr{\operatorname{tr}}
\def\sign{\operatorname{sgn}}

\def\fstp{\,\mbox{.}}

\def\be{\begin{equation}}
\def\ee{\end{equation}}
\def\nn{\nonumber}

\def\lag{\mathcal{L}}

\begin{document}

\maketitle


\tikzset{fixedpoint/.style={draw,circle}}
\tikzset{parameter/.style={gray}}
\tikzset{flow/.style={postaction=decorate,decoration={ 
			markings,%
			mark=at position .5  with {\arrow[line width=2pt]{>}}}}}


\clearpage
\section{Introduction}

One of the key features of non-relativistic theories is that they have a conserved particle number, and a complete lack of antiparticles. As a result, we can fix the particle number, and think of the theory as describing $N$-particle quantum mechanics. This means that interesting field theoretic phenomena must have direct equivalents in many-body quantum mechanics. In this paper, we will investigate this correspondence in the context of \textit{anyons} in 2+1 dimensions.

The quantum mechanics of interacting anyons is a famously rich problem, and apparently simple aspects such as the spectrum of anyons in a harmonic trap are surprisingly poorly understood. Part of the problem is that anyons are inherently strongly interacting; perturbative treatments are possible around the bosonic and fermionic theories, but otherwise analytic results are difficult to obtain. There are certain exceptions, such as the so-called ``linear states'' in a harmonic trap for special choices of contact interactions, whose energies are linear functions of the statistical parameter \cite{Chou:1991rg,Murthy:1992zm}. These also correspond to a family of operators in a field theory of anyons whose dimensions are exactly known \cite{cpt-anyons,cpt-nonanyon}, a fact which we will make use of here.

As we will explain in this extended introduction, these operators reveal that there is something special about theories of anyons experiencing attractive contact interactions. Specifically, if the interactions are tuned to create a two-anyon bound state at threshold, then sufficiently large numbers of anyons can consistently form bound states of arbitrary size or -- more dramatically -- collapse into point-like bound states. In fact, near the bosonic limit, the number of anyons this takes grows to infinity. This is the same limit in which mean-field theory is useful, and so ultimately the quantum behaviour impacts the \textit{classical} limit of the corresponding field theory.

The way this happens is more than a little surprising. It turns out that the relevant field theory supports solitonic states with finite, fixed particle number $N$. Unlike in relativistic contexts, where solitons are typically coherent superpositions of many-particle states in terms of fundamental quanta, here, they are directly linked to individual states in the underlying $N$-body quantum mechanics. In order to explain the finer points of this correspondence, we will stress that the field theory of interest has some surprising relevant and marginal operators which must be carefully accounted for.

In the rest of this introduction, we will set the scene with the relevant facts about anyonic theories, quantum and classical, before moving on to explain how these can be reconciled in section \ref{manybody}. A short summary of the paper is given in section \ref{summary}.

\subsection{Two Anyons}

Consider the relative wavefunction $\tilde{\psi}(r,\theta)$ of two Abelian anyons with statistical parameter $\nu=1/k$. By definition, the wavefunction should obey
\be
\tilde{\psi}(r,\theta+2\pi) = e^{2\pi i/k} \tilde{\psi}(r,\theta)
\ee
so that it is convenient to instead define a bosonic wavefunction $\psi = e^{-i\theta/k} \tilde{\psi}$. It follows that the action of the usual `free' Hamiltonian on $\psi$ is given by
\be
H \psi = \frac{1}{2m} \left[ -\frac{\partial^2}{\partial r^2} - \frac{1}{r} \frac{\partial}{\partial r} + \frac{(i\partial_\theta - \frac{1}{k})^2}{r^2} \right] \psi
\ee
i.e. the effective angular momentum is shifted.

However, this is not a complete description of the physics of two anyons; we must also supply a boundary condition at $r=0$. In order to give a well-defined theory, $H$ must be self-adjoint acting on the space of wavefunctions obeying this boundary condition, defining a so-called \textit{self-adjoint extension} of $H$.

Assuming for now that $k>1$, there is a one-parameter family of valid boundary conditions, affecting how the angle-independent `s-wave' states $\psi = \psi(r)$ behave as the particles approach each other \cite{manuel}. We find that we can consistently impose
\be
\psi(r) \sim C\left( \alpha \, r^{1/k} + r^{-1/k} \right)  \qquad \mbox{as $r \to 0$}
\ee
for any $\alpha$. This parameter is explicitly dimensionful, and it is convenient to instead write
$\alpha = -\mathrm{sign}(R_0) |R_0|^{-2/k}$
in terms of a \textit{signed} length $R_0$. 

A short calculation shows that for the case $R_0 > 0$, there is a bound state of energy $E=-E_0$ where
\be
E_0 = \frac{4}{mR_0^2} \left( \frac{\Gamma(1+1/k)}{\Gamma(1-1/k)} \right)^{k}
\label{anyonboundstateenergy}
\ee
which is given by a Bessel function $K_{1/k}(\sqrt{mE_0}r)$. On the other hand, for $R_0 < 0$ there is no bound state. We can think of these as distinguishing attractive and repulsive interactions respectively.

Note that there are two special points at which no dimensionful parameter is needed to describe the theory: $R_0 = 0 $ and $R_0 = \pm \infty$, corresponding to boundary conditions $\psi \sim r^{1/k}$ and $\psi \sim r^{-1/k}$ respectively. At these points, the theory is scale-invariant, which imbues it with a lot of extra structure, namely the $SO(2,1)$ symmetry of conformal quantum mechanics \cite{roman}. With either choice, there are linear states for many-body systems of anyons. The latter fixed point is especially interesting, as it describes a resonant interaction of the anyons -- there is a bound state tuned exactly to threshold, so that it has binding energy $E_0=0$ and infinite spatial extent.

As an aside, something slightly different happens for bosons in the limit $k\to\infty$, where $r^{\pm 1/k}$ coincide. The valid self-adjoint extensions are
\be
\psi(r) \sim C \left( \beta - \log \frac{r}{R_0} \right)
\label{bosonicbcs}\ee
where
$\beta=\log 2-\gamma_E$%
. This time, there is only one scale-invariant choice, as $R_0 = 0$ and $R_0 = \infty$ are equivalent boundary conditions. Moreover, there is always a bound state $\psi(r) \propto K_0(r/R_0)$ for any non-trivial $R_0$.

\subsection{The Field Theory of Abelian Anyons}

We are interested in the field theoretic formulation of the above analysis \cite{jpi,nishidason}. The central ingredient in anyonic theories in 2+1 dimensions is typically a Chern-Simons term. Indeed, a $U(1)_k$ Chern-Simons theory coupled to a single non-relativistic scalar field with Lagrangian and Hamiltonian
\begin{align}
\lag = \qquad &\int \rmd^2 x \  -\frac{k}{4\pi} a \wedge \rmd a + i \phi^\dagger D_t \phi
- H \label{anyonlag}
\\
H = \frac{1}{2m} &\int \rmd^2 x \  |D_i \phi|^2 + \mu |\phi|^2 + \lambda |\phi|^4 + \cdots 
\nn
\end{align}
reproduces the physics of Abelian anyons discussed above; we will look at non-Abelian anyons later, in section \ref{nonabelian}. Here, the index $i=1,2$ runs over spatial indices. 

We have, somewhat na\"ively as we shall see, adopted the approach of writing an effective field theory containing the operators of lowest engineering dimensions. For a non-relativistic scalar field, $\phi$ has dimension 1. The kinetic term has dimension 4, which is marginal in non-relativistic field theories, once dressed with the mass $m$ of $\phi$ excitations. In general, it is natural to take out an overall factor of $m$ from the Hamiltonian, and we have done so above.

The term $|\phi|^2$ has dimension 2 and is relevant. However, the global $U(1)$ symmetry guarantees that
\be
N = \int \rmd^2x \ |\phi|^2
\ee
is a conserved quantity, namely \textit{particle number}. Hence $\mu$ is a chemical potential, serving only to shift the energies of all states by an amount proportional to $N$. 
We will fix $N$, and think about this non-relativistic field theory in terms of the quantum mechanics of $N$ particles by working in a basis of states
\be
\left| \mathbf{x}_1, \ldots, \mathbf{x}_N \right> = \phi^\dagger(\mathbf{x}_1) \cdots \phi^\dagger(\mathbf{x}_N) \left| 0 \right>
\qquad \implies \qquad
|\phi|^2(\mathbf{x}) = \sum_{i=1}^N \delta(\mathbf{x}-\mathbf{x}_i)
\ee
and so we can safely set $\mu = 0$. (The physical states are supplemented with Wilson lines.)

In our $N$ particle quantum mechanics, the $|\phi|^4$ interaction provides a pairwise delta function interaction between the particles. But in two spatial dimensions, the delta function interaction of quantum mechanics requires regularization -- all it can do is modify the boundary conditions on wavefunctions as they approach each other, and in the previous section we found that this generically introduces a dimensionful quantity. Therefore, $\lambda$ undergoes dimensionful transmutation; for example, bosons experiencing a regularized, square potential of width $a$ have the boundary condition \eqref{bosonicbcs} with the physical parameter $R_0 = a e^{-2\pi/\lambda}$.

Our goal is to understand phenomena both in the language of field theory and quantum mechanics, however. Therefore, we turn to the RG analysis of the theory \eqref{anyonlag} \cite{bergman}. One finds there are generically logarithmic divergences, so that $\lambda$ flows non-trivially, but that there are two possible fixed points of the flow, with
\be \lambda = \pm \frac{2\pi}{k} \fstp \label{anyonfps} \ee
At these points, we actually have a fully fledged non-relativistic conformal field theory \cite{nishidason}.

One could also ask what the flow of the $\lambda$ coefficient looks like in each of these theories. The analysis is fairly trivial, as $|R_0|$ decreases along the flow.
This shows that $R_0 = 0$ is an stable fixed point, whilst $R_0 = \pm\infty$ is unstable. This is expanded upon a little in Appendix \ref{apprg}.

In this paper, we are going to focus on the unstable fixed point, where $\psi \sim r^{-1/k}$ is purely attractive, so $\lambda = - \frac{2\pi}{k}$. After all, many-body phenomena are more interesting if particles approach each other closely, as this gives them an opportunity to interact. This only requires a single parameter to be tuned, and working at the conformal point gives us a lot of extra tools with which to study what is, in general, a complicated interacting system. Moreover, as mentioned above, this fixed point corresponds to tuning a two-anyon bound state to threshold; theories with such resonances can play host to a variety of interesting phenomena.

\subsubsection{Jackiw-Pi Vortices}

The first startling observation about the $\lambda < 0$ fixed point is that the classical equations of motion have a moduli space of non-trivial zero-energy solutions \cite{jpvort,jpi}.

To find these solutions, we use Gauss's law to rewrite the Hamiltonian as
\be 
H = \frac{2}{m} \int \rmd^2 x \  |D_{\bar{z}} \phi|^2
\ee
and then solve the first-order equation
\be 
D_{\bar{z}} \phi = 0 \label{jpfirstorder}
\ee
together with Gauss's law. Solving \eqref{jpfirstorder} for the gauge field $a$ and substituting into Gauss's law reduces the equations of motion to Liouville's equation
\be
\nabla^2 \log |\phi|^2 = - \frac{4\pi}{k} |\phi|^2
\ee
which has infinitely many solutions. These take the form
\be
|\phi|^2 = \frac{2k}{\pi} \left| \frac{\partial_z f}{1+|f|^2} \right|^2 \label{generaljpsol}
\ee
where $f(z)$ is a holomorphic function of the coordinate $z = x +iy$. Focussing on those with finite particle number $N$, one finds that this is in fact quantized as 
\be
N = \int \rmd ^2 x \ |\phi|^2 = 2k q \qquad \text{for $q =0,1,2,\ldots$}
\ee
and moreover there is a convenient parametrization of generic configurations 
\be
f(z) = \sum_{i=1}^q \frac{c_i}{z-z_i}
\ee
where $z_i,c_i \in \mathbb{C}$ respectively describe the position, size and phase of $q$ separated vortices. These are known as \textit{Jackiw-Pi vortices}.

\begin{figure}[H]
	\begin{tikzpicture}
	\begin{axis}[domain=-1:1,y domain=-1:1,ticks=none]
	\addplot3[surf] {1/(0.2+(x*x+y*y))^2};
	\end{axis}
	\end{tikzpicture}
	\caption{A single Jackiw-Pi vortex.}\label{fig:onejp}
\end{figure}
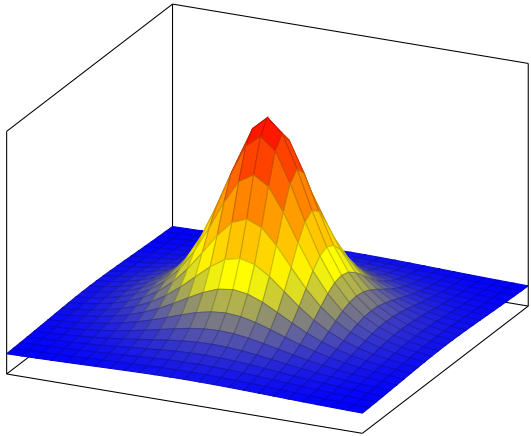

One surprise is that their magnetic charge of $2q$ is even, rather than merely an integer as required by charge quantization. As we allude to later on, this can possibly be explained in relation to the \textit{fusion rules} of $U(1)_k$ (or more prosaically the fact that, in $U(1)_k$, objects with odd flux are fermions if $k$ is odd). But perhaps the most striking thing about these configurations is that, unlike the case of topological defects which minimize the energy subject to some global constraint, they are \textit{non-topological}. As such, there does not appear to be any obvious reason for such a classical solution to exist at all.

In \cite{cpt-anyons}, it was suggested that their existence is linked to the behaviour of certain operators in the quantum mechanical theory. In the next section, we will explain this idea. In section \ref{manybody}, we will flesh it out.

\subsubsection{Operators at Unitarity}

The second interesting property of our chosen fixed point comes from performing a study of operator dimensions. In general, this is a hard problem which can only be tackled by e.g. an order-by-order expansion in $1/k$ around the bosonic theory.

However, there is a particular class of operator whose dimensions can be computed exactly. The expansion only has a term of order $1/k$, leading to them often being referred to as \textit{linear}.\footnote{One way to understand this is that the theory enjoys a hidden supersymmetry if one adds in a fermionic field. (This does not affect the purely bosonic sector at all in a non-relativistic theory, as there are no antiparticles and hence no loops involving fermions.) Then there are certain \textit{BPS operators} preserving enough supersymmetry that there are only one-loop corrections in $1/k$ \cite{cpt-anyons,cpt-nonanyon} -- the linear operators.}
The simplest such operators are $\phi^{N}$, which apparently have the dimension
\be \Delta_{N} = N - \frac{N(N-1)}{2k} \label{phindim} \ee
with the most general operator also containing $\ell$ holomorphic derivatives $\partial_z$, and having a dimension which is larger by $\ell$.
The minus sign in this expression means that as $N$ grows, $\Delta_{N}$ starts to decrease without bound. This has two important consequences.

The first is that na\"ively including these operators will ultimately violate the bound $\Delta \ge 1$ which is required for the theory to be unitary \cite{nagain}. More specifically, we find that
\be
\Delta_{2k} = 1
\ee
and $\Delta_{N} < 1$ for all larger $N$. This means that the latter operators should not be na\"ively included in the correctly quantized theory, whilst the operator $\phi^{\dagger 2k}$ behaves like a free field creation operator: in the Heisenberg picture, one can prove that any operator at the unitary bound satisfies the free Schr\"odinger equation. In \cite{cpt-anyons}, it was proposed that it is not a coincidence that this happens at the same value of the particle number $N=2k$ at which the Jackiw-Pi vortex appears. It seems reasonable to speculate that $\phi^{2k}$ indeed creates the Jackiw-Pi vortex. However, this leaves some puzzles. In particular, what is the interpretation of the size modulus of the Jackiw-Pi vortex?

The second consequence, not explored in \cite{cpt-anyons}, is that other operators which are na\"ively irrelevant can actually become relevant, affecting the universal behaviour of the theory. Consider gauge-invariant operators $|\phi^N|^2$ of dimension $2\Delta_{N}$. We know that relevance requires
\be
1 \le 2\Delta_{N} < 4 \fstp
\ee
For $k>3$, we find that $|\phi|^2$ and $|\phi|^{4k}$ have dimension $2$ like chemical potential terms, whilst $|\phi|^4$ and $|\phi|^{4k-2}$ both have dimension $4 - 2/k$ making them narrowly relevant at large $k$. All the other operators are irrelevant.%
\footnote{The situation is actually the same for $k=2$, whilst for $k=1$ only $|\phi|^2,|\phi|^4$ are distinct operators and both have $\Delta=1$. The case of $k=3$ is different, as the four relevant operators above are supplemented by the  marginal operators $|\phi|^6$ and $|\phi|^{8}$. We will look at small values of $k$ in section \ref{smallksec}.}

\subsection{Plan of the Paper and Key Results}\label{summary}

In this introduction, we have reviewed a standard story about two-body local interactions of anyons in two spatial dimensions. We have recapped how dimensional transmutation is realized in quantum mechanics, and understood RG fixed points both in the language of quantum mechanics and field theory. We have also introduced a theory with some particularly interesting properties -- non-topological vortices, and large operators which have very low dimension.

In section \ref{manybody}, we will extend our discussion of two-body physics to many-body systems. We will show that the presence of the relevant operator $|\phi|^{4k-2}$ signals that we have implicitly tuned another, many-body bound-state to threshold in assuming the formula for linear states is valid for large particle number. Generically, there is a new scale which enters the theory and changes the physics of systems with $2k-1$ or more particles. In particular, it alters the physics of systems of $2k$ particles, affecting the operator $\phi^{2k}$ which used to sit at unitarity.

We will then argue that the Jackiw-Pi vortex, whilst intimately related to the $\phi^{2k}$ operator which saturates the unitarity bound when we tune the interaction to threshold, can only be properly understood by allowing ourselves to move slightly away from this fixed point. In particular, we claim that the previously mentioned relevant operator explains the existence of the Jackiw-Pi size modulus, and that in the large $k$ limit this vortex looks essentially identical to a quantum many-body state whose size is set by this interaction. We will also offer some comments on states with more than $2k$ particles, or multiple Jackiw-Pi vortices, as well as on what happens for smaller values of $k$.

Finally, in section \ref{nonabelian}, we will look at some more complicated theories of non-Abelian anyons with similar behaviour. We will note that the theory of a single adjoint field is particularly natural, and behaves very similarly to the Abelian theory: the classical solutions are known, and the operators which lead to unitarity violation match them perfectly.


\clearpage
\section{Many-Body Physics and Jackiw-Pi Vortices}\label{manybody}

So far, we have essentially obtained an understanding of the $|\phi|^4$ interaction by thinking about two-particle physics. But the interesting phenomena described in the introduction affect $N$ on the order of $2k$, which is large in the classical limit where we might expect to gain an understanding of the Jackiw-Pi vortices. Hence we need to understand what happens as we crank up the number of particles.

We will stick to working in the anyonic theory \eqref{anyonlag} and tune our $\lambda$ coupling to the unstable fixed point. This means that the particles experience an attractive interaction in a scale-invariant manner, and whenever we bring two particles together we expect the wavefunction to behave as
\be \psi(\mathbf{x}_1,\ldots,\mathbf{x}_N) \sim |\mathbf{x}_i - \mathbf{x}_j|^{-1/k} \fstp \ee
In the introduction, we stressed two key consequences of working at this fixed point. Firstly, we pointed out that when we worked at high particle number, the operators ceased to be well-behaved, and one of them looked like it would create a free particle. Secondly, we saw that (for $k>3$ at least) we would find an additional relevant operator which would generically affect the physics of large numbers of particles. We will build an understanding of the quantum mechanical version of these observations in section \ref{manybodysubsec}, after exploring the two-body interactions a little more.

\subsection{More on Two-Body Physics}\label{moretwobody}

We should start by translating what we have learned about operator dimensions into the language of quantum mechanics. In particular, let us understand the dimension of the operator $\phi^2 (\mathbf{x}_1)$, obtained as the most singular term in the operator product expansion (OPE) of $\phi(\mathbf{x}_1)\phi(\mathbf{x}_2)$ as we take $\mathbf{x}_1, \mathbf{x}_2 \to \mathbf{x}$. From \eqref{phindim}, we know this has dimension $\Delta_{\phi^2} = 2 - \frac{1}{k}$, whilst $\Delta_\phi = 1$. Hence
\be
\phi(\mathbf{x}_1)\phi(\mathbf{x}_2) \sim |\mathbf{x}_1 - \mathbf{x}_2|^{-1/k} \phi^2(\mathbf{x})
\ee
But we can also think of $\phi^{2\dagger}$ as creating an interesting two-particle state, and ask how its wavefunction behaves as $\mathbf{x}_1, \mathbf{x}_2 \to \mathbf{x}$. We find that
\begin{align}
\psi(\mathbf{x}_1,\mathbf{x}_2)&= \left<0\right| \phi(\mathbf{x}_1)\phi(\mathbf{x}_2) \int \rmd^2 \mathbf{y} \ \psi_{\rm COM}(\mathbf{y}) \phi^{2\dagger}(\mathbf{y}) \left|0\right>
\\
&\sim  |\mathbf{x}_1 - \mathbf{x}_2|^{-1/k}  \int \rmd^2 \mathbf{y} \ \psi_{\rm COM}(\mathbf{y})  \left<0\right|  \phi^2(\mathbf{x})\phi^{2\dagger}(\mathbf{y}) \left|0\right>
\\
&=|\mathbf{x}_1 - \mathbf{x}_2|^{-1/k}\psi_{\rm COM}(\mathbf{x})
\end{align}
so the operator dimension directly translates into a boundary condition on the two-body wavefunction as the particles approach each other.

In fact, in this theory,
\be
\psi(\mathbf{x}_1,\mathbf{x}_2) = |\mathbf{x}_1 - \mathbf{x}_2|^{-1/k}
\ee
is annihilated by the Hamiltonian and is a valid two-body wavefunction at zero energy, which we know is interpreted as a bound state at threshold.
(By contrast, at the other fixed point, the two-body interaction is irrelevant, with $\Delta_{\phi^2} = 2 +\frac{1}{k}$, and the ground state is $\psi = |\mathbf{x}_1 - \mathbf{x}_2|^{+1/k}$, which should be interpreted instead as a scattering state.)

Given that a wavefunction with these asymptotics diverges at short distances, however, we should check the UV behaviour is well-defined. We find that the UV contribution scales as
\be
\int \rmd^4 \mathbf{x} \ |\psi|^2 \sim \int \rmd^2 \mathbf{x}_{\rm COM} \int r \, \rmd r \ r^{-2/k}
\ee
which is UV finite for $k>1$. 
Thus this is a perfectly sensible boundary condition to impose.

\subsubsection*{Fermions and Dimers at $k=1$}

It will prove useful to understand the limiting case of $k=1$, however. Something special happens here: the statistical phase is exactly $\pi$, and our anyons become fermionic. In two dimensions, we typically insist that fermions cannot have contact interactions, precisely because the only normalizable boundary condition compatible with self-adjointness is that the wavefunction vanishes as $\psi \propto r$. 
By contrast, the unstable fixed point requires the two-body boundary condition $\psi \propto 1/r$, which leads to a logarithmically non-normalizable wavefunction.

Nonetheless, there is a way to give meaning to this alternative attractive fixed point. The non-normalizability of the wavefunction arises at $r=0$, where the two particles are localized to the same point in space. Thus we can think of the divergent wavefunction as having all of its weight concentrated at $r=0$. More rigorously, we could introduce a UV length scale $a$ to regularize the theory, and then seek a wavefunction which had the above asymptotics down to this length scale. The probability mass of the wavefunction would then be localized on the length scale $a$. If we then removed the regulator, the correct conclusion would be that there was a point-like two-particle state, or \textit{dimer}, in the two-particle spectrum.\footnote{In fact, at $k=1$, we can independently form dimer states in either or both of the $\ell = \pm 1$ angular momentum channels, due to the parity symmetry of a fermionic theory. Since our interest is in parity non-invariant theories at general $k$, we will focus on having a single dimer.}

\begin{figure}[H]
\begin{tikzpicture}[scale=1]
\draw[domain=0.2:3,smooth,variable=\x,blue] plot ({\x},{0.1/(\x*\x)});
\draw[domain=-3:-0.2,smooth,variable=\x,blue] plot ({\x},{0.1/(\x*\x)});
\draw[domain=-0.2:0.2,smooth,variable=\x,green] plot ({\x},{62.5*\x*\x});
\draw[opacity=0.3,pattern=north west lines, pattern color=red] (0,0.2) ellipse (1.2cm and 0.5cm);
\draw[<->] (-1.2,-0.5) to node[below] {$\sim a$} (1.2,-0.5) ;
\end{tikzpicture}
\caption{The probability mass of the wavefunction $|\psi|^2$ of a dimer is localized on a UV length scale $a$. The limit $a \to 0$ may be taken consistently for these states as they obey the unitarity bound $\Delta \ge 1$ and therefore have positive norm. Their wavefunction becomes a delta function.}
\end{figure}
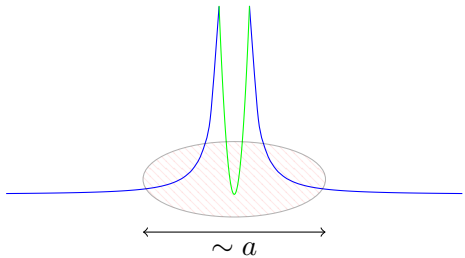

This is consistent with unitarity when we have this mild, logarithmic divergence of the wavefunction, which can be seen by observing that the dimension of the operator creating the dimer is $\Delta = 2-1/k = 1$ for fermions, which saturates the unitarity bound $\Delta \ge 1$.

The energy of the dimer state is a physical quantity to be determined by UV physics, a fact corresponding to the existence of the dimension 2 operator $|\phi|^4$ in the field theory, which acts as a chemical potential for the dimer. The unstable fixed point corresponds to giving this dimer zero energy; at the stable fixed point, it has infinite energy and decouples completely from the theory. In any case, the rest of the two-particle spectrum is simply that of a free fermion, as briefly discussed in Appendix \ref{apprg}.

A more common way of formulating an effective field theory with the dimer state would be to introduce an additional free field $\Phi$ representing the gauge-invariant dimer \cite{Nishida:2007de}, with Lagrangian
\be
\lag_\Phi = \int \rmd^2x \ i \Phi^\dagger \partial_t \Phi - \frac{1}{4m}\left(|\partial_i\Phi|^2 + \mu' |\Phi|^2 \right) \fstp
\ee
%
The new field has particle number $2$ and mass $2m$, and has bosonic statistics. Instead of tuning the $|\phi|^4$ interaction, we now tune the chemical potential to $\mu' = 0$ to reach a fixed point. We will return to this theory below.

\subsection{Many-Body Physics}\label{manybodysubsec}

Now let us return to general $k$, and consider what happens when we bring many particles together. If we only have the $|\phi|^4$ interaction present, then in fact
\be
\psi_0(\mathbf{x}_1,\ldots,\mathbf{x}_N) := \prod_{i<j} |\mathbf{x}_i - \mathbf{x}_j|^{-1/k}
\label{psi0}
\ee
is a valid many-body wavefunction (created using $\phi^{N}$), and also has exactly zero energy. The most singular UV behaviour is given by taking all particle separations of order $r$, where we find
\be
\int \rmd^{2N} \mathbf{x} \ |\psi|^2 
\sim \int \rmd^2 \mathbf{x}_{\rm COM} \int \rmd r \ r^{-N(N-1)/k} \cdot r^{2N-3} 
\sim \int \rmd r \ r^{2\Delta_N-3}
\ee
which is normalizable precisely when $\Delta_N > 1$.

\subsubsection*{Allowed Operators}

It is a general principle that operators with $\Delta \le 1$ create non-normalizable states \cite{nishy,cpt-anyons}. As we saw in the last section, non-normalizability does not generally preclude a sensible interpretation of the resulting operator -- if $\Delta = 1$, the operator is simply a free field operator, corresponding to a point-like many-particle state. However, if $\Delta < 1$ then the operator necessarily creates a state of negative norm, violating unitarity.

So for the anyonic theory, where we find $\Delta_{2k} = 1$, the natural interpretation is that $\phi^{2k}$ creates a free particle. However, the na\"ive result $\Delta_{N} < 1$ for $N > 2k$ signals that certain other operators must be taken out of the spectrum. For example, $\phi^{2k+1}$ does not create a valid state. Neither does $\phi^{2k} \partial_z \phi$, as this is a total derivative of the former operator and so not primary; from the point of view of the wavefunction, it differs only in the behaviour of the centre of mass.

But $\phi^{2k}\partial_z^2 \phi$ corresponds to a valid operator, in which the additional particle is given orbital angular momentum $\ell=2$ relative to the other particles. It has dimension $\Delta = 2$. (Strictly speaking this operator mixes with operators where the derivatives act differently; it should be understood throughout that we mean a particular choice which is a primary operator. In this instance, it would be $\phi^{2k}\partial_z^2 \phi - \phi^{2k-1}(\partial_z \phi)^2$, which corresponds to a factor in the relative wavefunction of $\sum_{i<j}(z_i-z_j)^2$. If $z_1=\cdots=z_{2k}=Z$ are bound together, then this wavefunction reduces to $(z_{2k+1}-Z)^2$, hence the interpretation of giving the additional particle angular momentum $2$ relative to the bound state.)

One can repeat this construction, adding $m \le 2k$ further particles with angular momentum relative to the original $2k$ particles. The resulting lowest-dimension operator is 
\be
\phi^{2k}(\partial_z^2 \phi)^m \qquad  \mbox{with dimension} \qquad \Delta = 1 + m - \frac{m(m-1)}{2k} \fstp
\ee
To make sense of this result, we recall the approach of adding a new free field $\Phi \sim \phi^{2k}$ to the field theory. (This time, the field has mass $2k$ times larger than that of $\phi$.) In fact, the above formula generalizes to an operator with particle number $N=2kq+m$:
\be
\Phi^{q}\phi^m \qquad  \mbox{with dimension} \qquad \Delta = q + m - \frac{m(m-1)}{2k}
\ee
where the interpretation of this operator in the language of the original theory is that
\be
\Phi^{q}\phi^m \qquad \longleftrightarrow \qquad \phi^{2k}(\partial_z^2 \phi)^{2k} \cdots (\partial_z^{2q-2} \phi)^{2k} (\partial_z^{2q} \phi)^{m} \label{oldtheorylongoperator}
\ee
and one may indeed verify that the dimensions match whichever theory we calculate in. Note that each time we form a new $2k$-particle block, we are obliged to add the next operator with angular momentum relative to every existing $2k$ block. This accounts for the increasing number of derivatives we insert as $q$ grows.

The key point is that a correction enumeration of the operators at this fixed point is restricted to a special subset which can most easily be explained by insisting that $\phi^{2k} \equiv \Phi$ be replaced by a free field operator in the OPE. Note that this new field is bosonic, just as the dimer field for $k=1$, as the statistical parameter of a bound state of $N$ anyons is
\be \nu = \frac{N^2}{2k} = 2k \in \mathbb{Z} \fstp \ee
This reflects the fact that there is a local monopole operator in $U(1)_k$ Chern-Simons theory with electric charge $k$, which can screen the charge of any state with particle number a multiple of $k$. (For $k$ odd, note that the $N=k$ state would be fermionic, so to generically obtain a bosonic state we actually need $N=2k$ particles. See Appendix \ref{fusionrules} for some brief comments on fusion rules.) The energetic cost of such a monopole operator is of course set by UV physics, and once more there is a new chemical potential associated with the $|\Phi|^2$ operator (or, equivalently, the $|\phi|^{4k}$ operator) which has dimension $\Delta = 2$. Strictly, in the CFT, one requires the coefficient of this operator to be tuned to $0$, though it is a relatively trivial deformation of the theory, being a chemical potential corresponding to a new conserved quantity.

As an aside, we should note the connection of the form of the operators (as expressed only in terms $\phi$) to $2k$-clustered symmetric polynomials, and also to certain wavefunctions in the fractional quantum Hall effect.  We are seeking the lowest-dimension linear operator which is no worse than logarithmically divergent; we can assume it has definite angular momentum. Consider the relative wavefunction of $N$ particles, $\psi = f(z_1,\ldots,z_N) \prod_{i<j}|z_i-z_j|^{-1/k}$. We require $f$ to be a translationally invariant symmetric function of minimal degree such that when more than $2k$ of its arguments coincide, the wavefunction vanishes. This, by definition, means seeking the $2k$-clustered polynomial of minimal degree. All such polynomials can be expressed in terms of Jack polynomials, which are labelled by partitions $\lambda_1 \ge \lambda_2 \ge \cdots \ge \lambda_N$ satisfying $\lambda_{i} \ge \lambda_{i+2k} + 2$ \cite{jackclustering}. For $N=2kq+m$, one finds the minimal such partition is
\be \lambda = (\underbrace{2q,\ldots,2q}_m,\underbrace{2q-2,\ldots,2q-2}_{2k},\ldots,\underbrace{2,\ldots,2}_{2k},\underbrace{0,\ldots,0}_{2k}) \ee
which corresponds precisely to the operator appearing on the right-hand side of  \eqref{oldtheorylongoperator}. These functions arise in $\mathbb{Z}_{2k}$ parafermionic field theories and as Read-Rezayi quantum Hall states.

\subsubsection*{Relevant and Marginal Operators}

We can also learn something about more conventional relevant and marginal operators, which for us will be those of the form $|{\cal O}|^2$ where ${\cal O}$ has dimension $1 < \Delta \le 2$. Let us focus on $N$ particle operators $\phi^N$ for now. Define
\def\coord{u}
\be
\coord^2 = \sum_{i=1}^N |\mathbf{x}_i - \bar{\mathbf{x}}|^2 = \frac{1}{N} \sum_{i<j} |\mathbf{x}_i - \mathbf{x}_j|^2
\ee
to be a collective, \textit{hyperradial} coordinate vanishing only when all particles coincide at their centre of mass $\bar{\mathbf{x}} = \frac{1}{N}\sum_i \mathbf{x}_i$. Now consider the action of the Hamiltonian on a wavefunction of the special form
\be
\psi = \tilde{\psi}(\coord) \cdot \psi_0
\ee
depending only on this highly symmetric coordinate, and the simplest wavefunction $\psi_0$ defined in \eqref{psi0}. We find that
\be
H \psi = - \frac{1}{8m} \left( \frac{\rmd^2}{\rmd \coord^2} + \frac{2\Delta_N - 3}{\coord} \frac{\rmd}{\rmd \coord} \right) \tilde{\psi} \cdot \psi_0
\ee
so the Hamiltonian preserves this set of states, and moreover the eigenvalue problem reduces to the Bessel equation, just as for two particles. We can also now perform the same analysis of self-adjoint boundary conditions in this sector as we did for two particles in the lowest angular momentum mode. The result is that we have in principle a choice of boundary condition at $\coord = 0$,
\be
\psi \sim \left(A + B \coord^{2(2 - \Delta_N)} \right) \cdot \psi_0
\ee
with logarithms appearing instead if $\Delta_N = 2$. One may verify that for $1 < \Delta < 3$, both are normalizable. 

For relevant operators with $\Delta_N < 2$, the ratio $B/A$ has a positive dimension and grows under RG flow. They therefore indicate an instability in the trivial $B=0$ boundary condition, and the system wants to head instead towards $A=0$ under RG flow. (At this second fixed point, $\phi^N$ has the dimension $\Delta_N + 2(2 - \Delta_N ) = 4 - \Delta_N > 2$ and so $|\phi^N|^2$ is irrelevant.%
)

So as expected, when there is a relevant operator, there is an additional dimensionful parameter we are tuning from an EFT point of view to reach the conformal theory we have been studying. Alternatively, we could allow this operator to turn on and investigate how it affects the theory. As before, for one sign of this parameter, it determines the size and energy of an $N$-particle bound state whose wavefunction is given by a Bessel $K$ function of the $\coord$ coordinate:
\be
\psi_{\rm bound} \propto \coord^{2-\Delta_N} K_{\Delta_N - 2}\left(\sqrt{8mE_0'} \coord\right) \cdot \psi_0
\label{newbound}
\ee
Importantly, this means that the na\"ive fixed point around which we were originally working ($B=0$) should be thought of as describing a theory with an $N$ particle bound state tuned to threshold.

For our anyonic theory, in the generic $k>3$ case, we first observe such a relevant operator when we hit $N=2k-1$. This, in general, also affects the physics of $N \ge 2k$ particles, as the modified boundary condition must be enforced for each subset of $2k-1$ particles, which changes the spectrum substantially. Generically, this new boundary condition spoils the supersymmetric structure which allowed us to compute operator dimensions exactly.

Considering a theory with $\tilde{\Phi}$ representing the $(2k-1)$-particle state may make it easier to study the effects of modifying the above boundary condition, which would appear as a chemical potential for $\tilde{\Phi}$. The charge $2k$ operator would then arise as $\phi\tilde{\Phi}$. Though we will not pursue this, it might prove a productive way to try and better understand the phenomenology of these states. For our purposes, when this interesting relevant operator is turned on, the main result we will use is the possible existence of the bound state \eqref{newbound}. It seems difficult to perform any other exact computations.

When the $(2k-1)$-particle operator is tuned to the unstable fixed point, however, we can still compute dimensions  of the linear operators, and this allows us explicitly list all such operators which are not irrelevant. The set of allowed operators is given by \eqref{oldtheorylongoperator}, and these give precisely two more deformations that are not irrelevant (for $k>3$). These are most elegantly written as $|\Phi|^2|\phi|^2$ and $|\Phi|^4$, and both are marginal. This information is summarized in Table \ref{dimtable}.
\begin{table}[]%
	\makebox[\textwidth][c]{%
	\begin{tabular}{cccc}
		Operator & Dimension & Interpretation  \\\hline
		$|\phi|^2$ & $2$  & Chemical potential for $\phi$   \\
		$|\phi|^4$ & $4 - \frac{2}{k}$ & Pairwise anyon interaction, tuned to threshold  \\
		$|\phi|^{4k-2}$ & $4 - \frac{2}{k}$  & $2k-1$ anyon interaction, tuned to threshold \\
		$|\Phi|^2 \sim |\phi|^{4k}$ & $2$ & Chemical potential for tightly bound $2k$ anyon state \\
		$|\Phi|^2|\phi|^2 \sim |\phi|^{4k}|\partial_z^2\phi|^2$ & $4$ & Interaction of tightly bound $2k$ state with single anyon (marginal) \\
		$|\Phi|^4 \sim |\phi|^{4k}|\partial_z^2\phi|^{4k}$ & $4$ & Pairwise interaction for tightly bound $2k$ anyon state (marginal)
	\end{tabular}%
}%
\caption{Relevant and marginal linear operators at the supersymmetric fixed point for $k > 3$.} \label{dimtable}
\end{table}%

In summary, then: if we tune only the two-body interaction of anyons to threshold, then generically there will be a $(2k-1)$-particle bound state of arbitrary size and energy, breaking any supersymmetric structure in the theory. If we also tune this bound state to threshold, then we preserve supersymmetry and obtain a point-like $2k$-particle state which decouples from the rest of the theory.

\subsection{Quantization of Jackiw-Pi Vortices}

Now we want to understand how this relates to the Jackiw-Pi vortices mentioned in the introduction. We will focus on understanding a single Jackiw-Pi vortex centred at $z_0$, which is a classical field configuration with $N = 2k$ and the profile
\be
|\phi|^2 = \frac{2k}{\pi} \frac{|c|^2}{(|c|^2+|z-z_0|^2)^2} \fstp
\ee
where $c$ is the size modulus. It takes the form of a smooth bump with rotational symmetry about its maximum at $z=z_0$, as shown in Figure \ref{fig:onejp}.

Of course, whilst we have identified a quantum mechanical state with a free COM at $N=2k$ which we tentatively identified with a JP vortex, here we seem to have infinitely many states parametrized by $c$. To resolve this tension, we should quantize these vortices.

In order to do this, we should not simply perform the usual trick of promoting $z_0$ and $c$ to be functions of time before substituting into \eqref{anyonlag}. The reason is that the dynamics is first-order; the conjugate momenta are encoded directly in the field configuration rather than its time derivative.

However, we can use the symmetries of the problem to exactly compute, say, the Galilean boost of this configuration moving with complex velocity $\dot{z} = v$ \cite{Jackiw:1991ns}. This gives a configuration satisfying not $D_{\bar {z}}\phi = 0$ but instead
\be
D_{\bar{z}} \phi = \frac{i m v}{2} \phi \fstp
\ee
As a check, we can combine this with the $\phi^\dagger$ equation of motion to find
\be
D_{t}\phi = \frac{2i}{m}D_{z}D_{\bar{z}}\phi = -v D_{z} \phi
\ee
as expected. We can also derive
\be
\frac{\rmd}{\rmd t} \left<z\right>
= \frac{\rmd}{\rmd t} \frac{\int \rmd^2z \ z |\phi|^2}{\int \rmd^2z \ |\phi|^2}
= \frac{1}{N}{\int \rmd^2z \ z \left(\phi^\dagger (-v D_z) \phi + (-\bar{v} D_{\bar{z}}) \phi^\dagger \phi\right)} = v
\ee
using integration by parts.
More interestingly, we can evaluate the Hamiltonian,
\be
H = \frac{2}{m} \int \rmd^2z \ |D_{\bar{z}}\phi|^2
= \frac{1}{2} Nm|v|^2
\ee
as expected for an object of mass $Nm$ moving with velocity $v$. It follows that quantizing the translational mode of the vortex indeed leads to a free quantum mechanical particle of mass $Nm$.

Let us now turn to $c$. Firstly, the phase of $c$ corresponds solely to the fact that the field configuration breaks the global part of the gauge group, i.e. has non-zero particle number $N$. Under quantization, this simply leads to the existence of both a creation and destruction operator for the vortex as is usual for a particle, rather than contributing any dynamical modes.

More interestingly, the magnitude of $c$ affects the size of the classical JP vortex. What happens when we try to quantize this? There is in fact a classical way to arrive at a configuration with $c$ time-dependent: a conformal boost. One finds that a conformal boost with parameter $a$ has energy
\be
H = \frac{1}{2} Nm a^2 \left<|z|^2\right>
\ee
where
\be
 \left<|z|^2\right> =  \frac{\int \rmd^2z \ |z|^2 |\phi|^2}{\int \rmd^2z \ |\phi|^2}
\ee
diverges logarithmically at long distances for a single vortex since $|\phi|^2 \sim |z|^{-4}$. Hence we conclude that $c$ is \textit{not} in fact a normalizable, fluctuating mode of the vortex. Instead, we ought to think of the size modulus as being fixed when we quantize the system -- it is an extra dimensionful parameter.


The last observation we should make is that there are solutions with multiple Jackiw-Pi vortices. As mentioned previously, there is a whole moduli space consisting of configurations with Jackiw-Pi vortices of arbitrary size in arbitrary locations. When widely separated relative to the size moduli, these configurations look like simple superpositions of individual Jackiw-Pi vortices. We will not attempt to study this whole space here, though we anticipate from the above that at least one of these modes is not a fluctuating mode in the quantum mechanics.

\begin{figure}[H]
	\begin{tikzpicture}
	\begin{axis}[domain=-1:1,y domain=-1:1,ticks=none,fill opacity=0.2]
	\addplot3[surf,line width=0.25pt] {(x*x+y*y)/(0.12+(x*x+y*y)^2)^2};
	\end{axis}
	\end{tikzpicture}
	\caption{A pair of coincident Jackiw-Pi vortices in a rotationally symmetric state.}\label{fig:twojp}
\end{figure}
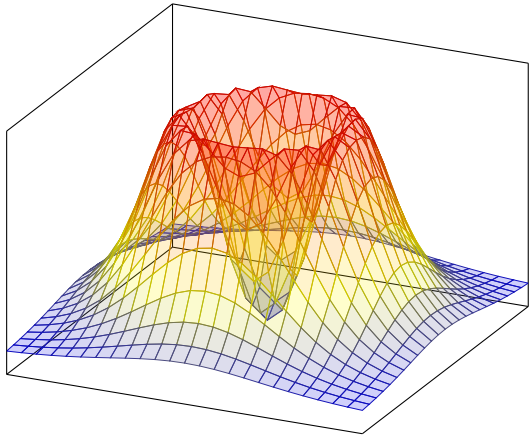

We will, however, note that something interesting happens if we insist that there are multiple Jackiw-Pi vortices sitting at one point in space. Imposing rotational symmetry and seeking solutions with $N = 2kq$ for $q>1$, one finds such configurations are given by \eqref{generaljpsol} for the choice $f = c^q/z^q$, and take the form
\be
|\phi|^2 = \frac{2k |c|^{2q}q^2}{\pi} \frac{ r^{2q-2}}{(r^{2q}+|c|^{2q})^2} \fstp
\ee
Rather than having a maximum at $r=0$ as in the $q=1$ case, this actually vanishes like $r^{2q-2}$ for $q>1$, indicating we cannot simply stack multiple vortices atop one another. The particular case of $q=2$ is shown in Figure \ref{fig:twojp}.

\subsection{The Correspondence}

We have seen that the quantum mechanics has an operator which becomes relevant and gives rise to a non-dynamical size modulus for the $N=2k-1$ state, and another operator which decouples as it hits a unitarity bound at $N=2k$. Meanwhile, quantizing the classical JP soliton gives a decoupled free particle with a non-dynamical size modulus at $N=2k$. The natural conclusion is that the two dimensionful moduli and the two free particles are both in direct correspondence.

In order to understand how this arises, consider first the CFT with the carefully tuned choice of $(2k-1)$-particle boundary condition. Then, fixing the centre of mass at the origin, there exists a single $(2k-1)$ state which has been tuned to threshold. If we add one more particle, it can bind to this state so tightly that the wavefunction has support only when all $2k$ particles are coincident, and this leads to a decoupled $\phi^{2k}$ operator at unitarity. This matches with the degenerate $c\to0$ limit of a Jackiw-Pi vortex of zero width, whose density profile is simple a delta function.

Now suppose we detune with a slight $|\phi^{2k-1}|^2$ interaction so as to create a shallow bound state for $2k-1$ particles. This bound state has a finite energy $E=-E_0'$, and a finite size in that $u \lessapprox 1/\sqrt{mE_0'}$, as can be seen from the exponential decay of the Bessel function in \eqref{newbound}.

However, one can readily compute that for large $k$, where we expect to make contact with the classical description, 
\be
\left<|\mathbf{x}_i|^2\right> = \frac{1}{2k-1}\left<u^2\right> \sim \frac{1}{kmE_0'}
\ee
is actually small for any fixed $E_0'$. Alternatively, we can take a limit $E_0' \to 0$ and $k\to\infty$, making contact with the classical limit of the na\"ive CFT, whilst holding this expectation value constant at
\be
\left<|\mathbf{x}_i|^2\right> = a^2 \fstp
\ee

Now we want to understand the behaviour of the $N=2k$ state arising from adding one more particle to the system. Specifically, we are interested in states which look approximately like the $(2k-1)$-particle bound state coupled to an additional new particle, with energy on the order of $E_0' \to 0$. (The state we are looking for should be smoothly connected to the state created by the $\phi^{2k}$ operator at the fixed point.)

When one particle is far separated from the centre of mass, we expect a good approximation to the wavefunction to be given by considering the bound state as a particle of charge $(2k-1)$ interacting through a Chern-Simons gauge field with a particle of charge $1$. This leads to a long-distance decay of the wavefunction 
\be
\psi \sim r^{-\frac{2k-1}{k}}
\ee
which, in the classical limit $k\to\infty$, gives a density that approximately behaves like $|\psi|^2 \sim r^{-4}$. This gives rise to the same long-range fall-off of the density $|\phi|^2$ observed for a Jackiw-Pi vortex configuration.

\begin{figure}[H]
	\begin{tikzpicture}
	\draw[opacity=0.3,pattern=north west lines, pattern color=red] (0,0.2) ellipse (1.2cm and 0.5cm);
	\draw[<->] (-1.2,-0.5) to node[below] {$\sim a$} (1.2,-0.5);
	\foreach \n in {(0,0),(0.2,-0.1),(0.5,0.2),(0.6,0.45),(-0.3,0.05),(-0.4,0.2),(-0.8,0.1),(0,0.4)}
	\node at \n [circle,fill,inner sep=1.5pt] {};
	\node at (4,0.2) [circle,fill=blue,inner sep=1.5pt] {};
	\draw[xshift=0cm,yshift=0.2cm,domain=0:5.5,smooth,variable=\x,blue] plot ({\x},{1.8/((1.3+\x*\x)*(1.3+\x*\x))});
	\node at (4,-0.5) {$|\psi|^2 \sim r^{-4 + \frac{2}{k}}$};
	\end{tikzpicture}
	\caption{Adding one particle to a $(2k-1)$-particle ground state of size $a$. At long distances $r \gg a$ from the centre of mass, it should be a good approximation to treat this as a charge $2k-1$ particle interacting with a charge $1$ particle.}\label{fig:extraparticle}
\end{figure}
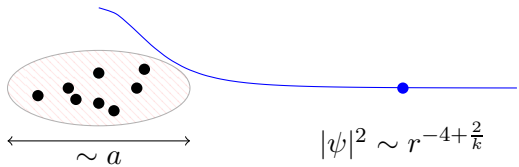

Meanwhile, near the centre of mass, we expect particles to be smeared fairly evenly over the length scale $a$ as they are only very weakly interacting at large $k$. It is not unreasonable to guess that one-particle density is approximately a constant set by $1/a^2$ near the centre-of-mass. This corresponds to the fact that Jackiw-Pi vortices have a smooth maximum attained at their centre of mass.

Note that the classical derivation of the Jackiw-Pi profile is essentially a mean-field calculation, wherein the gauge field is sourced by a continuous density profile rather than by point particles. So assuming that the mean-field description becomes a good approximations at large $k$, where $N=2k$ is also large, we would in fact expect that the profile of a JP vortex emerges in the classical limit.

Hence it seems that if we take the large $k$ limit with a particular choice of $|\phi^{2k-1}|^2$, we do indeed expect to find a single Jackiw-Pi-like state of some fixed size $a$.

We should take this opportunity to offer an explanation of why the Jackiw-Pi vortices should be quantized to have $2kq$ particles, and hence even $U(1)$ flux $2q$. We mentioned above that a $2k$ particle state is the lowest state which is guaranteed to be bosonic; in theories with odd $k$, states with odd $U(1)$ flux are fermions. This is presumably why the Jackiw-Pi vortex must have $N=2k$ -- it needs to correspond to a free bosonic excitation in the many-body quantum mechanics, even if we take $k\to \infty$ through only odd values of $k$. This is related to a conjecture made in Appendix \ref{fusionrules}.

\subsubsection*{Multi-Jackiw-Pi States}

The classical theory also has multi-Jackiw-Pi states of course. Back in section \ref{manybodysubsec}, we saw that at finite $k$, but with a $(2k-1)$-body interaction tuned so that $a=0$, the spectrum of the theory contains a free decoupled field representing the $2k$ particle state. This perfectly fits the interpretation of a classical moduli space of $2k$-particle states, with non-fluctuating size moduli all being taken to zero. It seems likely that at other values of $a$, and finite $k$, the corresponding states do interact non-trivially, but that their interaction energy goes to zero as $k\to\infty$ at fixed $a$, leaving a classical moduli space of zero-energy modes.

One simple observation which we can make is about what happens when we bring multiple Jackiw-Pi vortices together at one point. Classically, the configurations always have a zero, as depicted in Figure \ref{fig:twojp}. This suggests that the wavefunction of $\phi$ excitations vanishes when we try to bring multiple $2k$-particle states close to one another. This fits well with the observations of section \ref{manybodysubsec}, where we noted that we have to give $\phi$ excitations associated to different $2k$-particle states angular momentum relative to each other, forcing the wavefunction to vanish when they are brought too close.

We will not study these configurations in any more detail. We will simply note that considering new effective field theories (as proposed in section \ref{manybodysubsec}) may be a good way to study these states at finite $k$ in the presence of non-trivial interactions. There may also be an important role for the two marginal operators listed in Table \ref{dimtable}. One natural computation would be to study what happens if one quantizes the $\phi$ field around a non-trivial state containing one or two Jackiw-Pi vortices.

Additionally, it would be interesting to understand what happens when we add multiple flavours of matter. It follows from results in the section \ref{nonabelian} that there is a similar story to tell, as there are still vortices with total particle number quantized as $N=2kq$ as well as operators with this particle number which sit at unitarity. However, there are now additional moduli corresponding to the flavour degrees of freedom, and additional relevant and marginal parameters to consider.

\subsection{Small $k$ and Efimov Physics}\label{smallksec}

We have mainly focussed on the behaviour for these theories at large $k$, thinking of this as the classical limit. It is also, of course, the bosonic limit, in which the statistical phase $\pi/k \to 0$. But we have neglected the question of what precisely happens for small $k$.

We have already discussed, in section \ref{moretwobody}, the special fermionic theory arising at $k=1$. It is clear here that the operator at unitarity arising at $N=2k=2$ is necessarily a point-like excitation. (There is no additional relevant operator at $N=2k-1$ with which we can deform the theory.)

This is in fact an interesting system in itself, studied in \cite{Moroz:2013kf}. By introducing a new dimer field\footnote{In fact, they introduce operators for \textit{both} dimer operators, respecting the parity symmetry.} as seen in section \ref{manybodysubsec}, and investigating the behaviour of the three-body interaction, they argue that the fermions exhibit a so-called \textit{super Efimov effect}. This means that the three-body system has an infinite spectrum of bound states, spaced in energy according to a particular universal law. One way to demonstrate this is to find a limit cycle in the RG flow; in the case of the classic Efimov effect, one has a periodic limit cycle which leads to discrete scale invariance, and hence to infinite sequences of three-body bound states with a geometric distribution in energy. By contrast, the super Efimov effect arises in the non-generic case of a limit cycle which is \textit{homoclinic}, i.e. contains a fixed point, which spoils the periodicity of the trajectory. This leads to a more complicated but nonetheless still infinite spectrum of (in this case) three-body bound states along trajectories converging to the stable homoclinic cycle. One also finds four-particle resonances associated to each three-body state, as for the conventional Efimov effect.

It would be very interesting to see if this phenomenon occurs for $N=2k+1$ or $N=4k$ particles for other $k$. (It might even give a new role to the Jackiw-Pi size moduli, namely labelling a choice of state from the Efimov spectrum.) We have seen that the $(2k+1)$-body and $4k$-body interactions are both marginal at the fixed point, so it may indeed be that slightly detuning the theory leads to a limit cycle for these interactions, as we now explain.

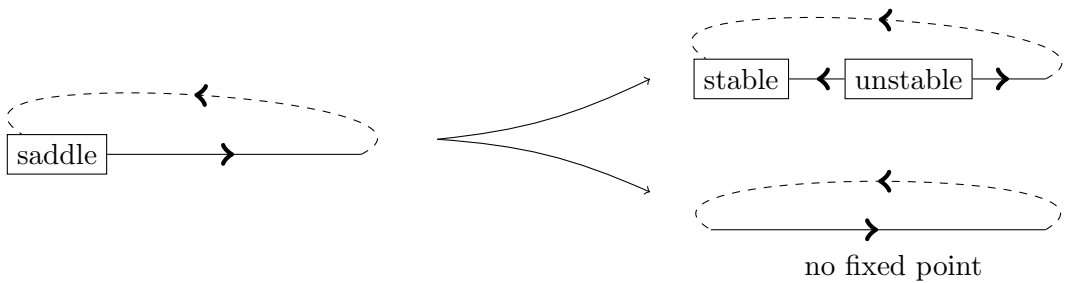
\begin{figure}[H]
\begin{tikzpicture}
\path[use as bounding box] (-3, -2.5) rectangle (11, 1);
\node[draw] (L) at (-3,-1) {saddle};
\coordinate (F) at (1,-1);
\draw[flow] (L) to (F);
\draw[flow,dashed] (F) to[bend right=150] (L);

\draw[->] (2,-0.8) to[bend right=10] (4.8,0);
\draw[->] (2,-0.8) to[bend left=10] (4.8,-1.5);

\node[draw] (L1) at (6,0) {stable};
\node[draw] (R1) at (8.2,0) {unstable};
\coordinate (F1) at (10,0);
\draw[flow] (R1) to (L1);
\draw[flow] (R1) to (F1);
\draw[flow,dashed] (F1) to[bend right=150] (L1);

\coordinate (F1) at (5.6,-2);
\coordinate (F2) at (10,-2);
\draw[flow] (F1) to (F2);
\draw[flow,dashed] (F2) to[bend right=150] (F1);
\node at (8,-2.5) {no fixed point};
\end{tikzpicture}
\caption{Two possible instabilities of saddle points associated to marginal operators in non-relativistic RG flow. Limit cycles are associated with infinite spectra of bound states.}\label{fig:splitsaddle}
\end{figure}

Suppose the flow due to a marginal deformation, like that of the two-body interaction in the bosonic theory, has a single fixed point being a saddle, which flows back to itself under large perturbations. Such a flow is typically unstable to deformations. One possible generic behaviour is that the saddle splits into two fixed points, one stable and one unstable; this is the behaviour of the two-particle interaction of anyons which we have seen has two fixed points. In that context, non-zero $1/k$ is thought of as a deformation from the bosonic theory. (See Figures \ref{fig:anyonrg} and \ref{fig:simplerg} in the Appendix.) Another possible behaviour is that the fixed point disappears, leaving no fixed points at all. Both are depicted in Figure \ref{fig:splitsaddle}. The latter is the most interesting case which gives rise to Efimov-like physics. If the perturbation is fixed under RG then we genuinely get a periodic flow; if it is instead irrelevant and itself flows to zero, so that we approach the homoclinic cycle containing the saddle, then we can find less generic behaviour as in the super Efimov effect.

We leave the question of the precise fate of these marginal couplings for future work.

To study other small values of $k$, it is helpful to turn to \cite{Nishida:2007de} which studies the problem of \textit{fermions} coupled to a Chern-Simons term introducing a small statistical phase. This is essentially perturbing around our $k=1$ theory, and can again be studied by introducing a dimer operator (although now it has a non-trivial anomalous dimension, rather than satisfying $\Delta = 1$). Near the free fermion point, a variety of interesting physical phenomena are manifested. At the level of engineering dimensions, there are two apparently marginal operators corresponding to three- and four-body interactions. The three-body interaction has two fixed points; tuning it to the unstable one, we find the four-body interaction undergoes a limit cycle, giving rise to an instance of the original Efimov effect \cite{efimov} -- an infinite sequence of bound states with geometrically spaced energy levels.

This analysis is expected to hold for statistical phases close to $k=1$. How does this connect to the picture we have been discussing? It turns out that the \textit{stable} choice of three-body interaction in \cite{Nishida:2007de} is identified with the theory we have been studying -- the formulae (30) and (31) in that paper agree with the dimensions of the operators \eqref{oldtheorylongoperator} which are expected to be permitted in our theory.\footnote{In fact, there is a typographical error in \cite{Nishida:2007de}, which the author has acknowledged: equation (30) should contain a term quadratic in $n$ identical to that in equation (31).} Hence we are not in the situation of expecting this type of Efimov effect in our theory near $k=1$. To see it, we would have to tune the three-body interaction.

It is interesting to note, however, that at $k=2$ we did tune an interaction of $2k-1=3$ particles! However, in this case we know that the four-body interaction simply determines the energy of a unique tightly bound four-particle state, so it seems unlikely that this type of Efimov physics is present here. This suggests that the results of \cite{Nishida:2007de} probably do not hold for $k=2$, though they may do so for statistical parameters corresponding to $1 < k < 2$.

However, there is an important caveat to all of the discussion above, which is that (at least near the fermionic point), there may extra relevant operators in the \textit{non-linear} sector. We have already mentioned one example of this, which is that free fermions can form \textit{two} distinct dimers with different angular momenta. One should certainly investigate this possibility in order to understand these theories completely.

Note that one might have wondered if tuning interactions for both $2$ and $2k-1$ particles to resonance would be enough to induce Efimov physics for $2k$ particles at large $k$. This appears not to be the case, however, as the theory has only a single bound state at this special point.


\clearpage
\section{Non-Abelian Theories}\label{nonabelian}
\def\rank{p}

So far, we have been focussed entirely on the Abelian Chern-Simons theory. However, various Jackiw-Pi-like configurations arise in non-Abelian gauge theories too. We will look mainly at $SU(\rank)_k$ Chern-Simons theories with $N_f$ scalar fields $\phi_a$, each transforming in some representation $R_a$ of the gauge group. The generators $t^\alpha[R_a]$ for $\alpha=1,\ldots,\rank^2-1$ are normalized such that in the fundamental representation $\tr [t^\alpha t^\beta] = \delta^{\alpha\beta}$. Then there is indeed a CFT of the form
\begin{align}
\lag = &\int \rmd^2x \ -\frac{k}{4\pi} \tr \left[a \wedge \rmd a - \frac{2i}{3} a\wedge a \wedge a\right] + i\phi^\dagger_a D_t \phi_a - H \\
H = \frac{1}{2m} &\int \rmd^2x \  \left|D_i\phi_a\right|^2 - \frac{2\pi}{k} \left(\phi^\dagger_a t^\alpha[R_a] \phi_a\right)\left(\phi^\dagger_b t^\alpha[R_b] \phi_b\right)\nn
\end{align}
which is again unstable to small quartic interactions for $k>0$. Similarly, a second, stable fixed point exists at the opposite sign of the quartic interaction \cite{bbak}. We could give the fields different masses, but for simplicity we have given them all mass $m$.
Note that a very similar construction is possible with product gauge groups, including the special case $U(\rank)$.

There are in principle many distinct quartic couplings we could add, arising from different irreps $R$ in the decomposition of $R_a \otimes R_b$. We have apparently made a very special choice here -- the reason is that once more certain exact computations can be done (and indeed the above theory has a hidden supersymmetry).

A related fact is that there is not a simple interpretation of the sign of the coupling as giving rise to attractive or repulsive interactions, because the particles can interact through distinct channels labelled by representations $R$ and may do so with different signs. Indeed, as can be seen in Section \ref{nonabopdimsec} below, our choice of interaction is generally attractive when the quadratic Casimir of $R$ is large (for example in the symmetric product) and repulsive when it is small (for example in the singlet).

One particularly interesting case we will look at involves a single complex boson $\phi$ transforming in the adjoint representation of $SU(\rank)$. In this case, there are typically four distinct quartic interactions. These arise from the decomposition
\be
{\rm Sym} \left[ \ \yng(2,1,1,1,1) \otimes \yng(2,1,1,1,1) \ \right] = \bullet + \yng(2,1,1,1,1) + \yng(2,2,1,1) + \yng(4,2,2,2,2) \label{adjointchannels}
\ee
shown here for the case $\rank = 6$.
The first channel involves the singlet $\tr [\phi^2]$; the second is an adjoint given by the traceless part of $\phi^2$; the third representation arises from $\phi^{[A}_{[B}\phi^{C]}_{D]}$; and the fourth from $\phi^{(A}_{(B}\phi^{C)}_{D)}$. (This holds for all $\rank \ge 4$. For $\rank=3$, we lose the third interaction. For $\rank=2$, we also lose the second one.) We will see below that at this fixed point, only the fourth, symmetrized operator is relevant, giving an attractive interaction. The other three are irrelevant, and the interaction is repulsive in these channels.\footnote{If one has two or more fields in the adjoint representation, one need not take the symmetric product as depicted in \eqref{adjointchannels}, and three more channels open up. One is irrelevant, but two prove marginal.}

Even more important than in the Abelian case is that fact that the representation theory of non-Abelian Chern-Simons theories is complicated by \textit{fusion rules}, certain selection rules arising from monopole operators in the theory. We will not address these in detail, though see Appendix \ref{fusionrules} for a brief discussion of their nature and significance -- it seems likely that there is something interesting to be said, and a simple conjecture is offered in the appendix.

\subsection{Non-Abelian Vortices}
Regardless of how many flavours of matter we include and what representations they transform under, the relevant equations are still a first-order equation and Gauss's law:
\begin{align}
D_{\bar{z}} \phi_a &= 0 \\
\frac{k}{2\pi} f^\alpha &= \sum_a \phi^\dagger_a t^\alpha[R_a] \phi_a \label{nonabgauss}
\end{align}
In some cases, these are well studied; most notably, the case with a single matter field transforming in the adjoint representation of $SU(\rank)$. In this case, the equations above are equivalent to the equations of \textit{unitons}, and the solutions are known \cite{Dunne:1992hq}. One finds that the quantization of energy of unitons translates into vortex solutions with quantized particle number
\be
N = kq  \qquad \text{for $q =0,1,2,\ldots$} \fstp
\ee

Whilst no such elegant classification is known for other matter content, large numbers of solutions can be more straightforwardly obtained even in the most general case \cite{Dunne:1990qe}. Suppose we turn on a single component $\phi_{a,\rho} = c_a \chi(z,\bar{z})$ corresponding to a single weight $\rho$ of a representation $R_a$, where we choose $\sum_a|c_a|^2 = 1$. This has the effect that the interaction terms $\phi^\dagger t^\alpha \phi$ are only non-trivial when $t^\alpha$ is an element of the Cartan subalgebra. It also suffices to consider turning on Cartan components of the gauge field.

Concretely, let $h^A$ span the Cartan subalgebra, and write $\mu^A$ for the weight associated to the weight vector $\rho$. Then the two equations above reduce to
\begin{align}
\partial_z \log \chi &= i\mu^A a_z^A \\
\frac{k}{2\pi} f^A &= \mu^A |\chi|^2
\end{align}
With the additional ansatz that $a_z^A \propto \mu^A$, we obtain simply
\be
\partial_{\bar{z}}\partial_{z} \log |\chi|^2 = - \frac{\pi\left<\mu,\mu\right>}{k} |\chi|^2
\ee
which is identical to the Liouville equation for the Abelian case, save for the appearance of the norm of the weight vector. Accordingly, the solutions obey the new quantization condition
\be
N = \frac{2kq}{\left<\mu,\mu\right>}  \qquad \text{for $q =0,1,2,\ldots$} \fstp
\label{simplenonabjpquant}
\ee

For the case of a single particle in the adjoint representation, where we have a classification of vortex solutions with particle number always a multiple of $k$, note the highest weight is a root satisfying $\left<\mu,\mu\right>=2$. Therefore the above analysis also predicts a (sub)set of solutions with precisely this quantization.
This formula, which is unchanged for the case of $U(1)_k$, also reproduces our earlier results if we take $\mu=1$, as appropriate for a scalar of unit charge.

This formula does not necessarily give an integer value for $N$. This is one reason why it seems like the adjoint representation, for which we do automatically obtain an integer, is a particularly natural case to study. Of course, for any $\mu$, one can simply choose $k$ such that $2k/\left<\mu,\mu\right>$ is an integer. Otherwise, one should exclude any such states from the spectrum of the quantum mechanical theory.


\subsection{Operator Dimensions}\label{nonabopdimsec}

As discussed in \cite{cpt-nonanyon}, there are still linear operators in this theory: those built entirely from $\phi_a$ and holomorphic derivatives. Suppose we choose an operator $\cO$ built from $N$ fundamental fields $\phi^\dagger_{a_i}$. It is helpful to decompose the tensor product $\bigotimes_{i=1}^N R_{a_i}$ into irreducible representations, and choose $\cO$ to transform in some particular irrep $R$. Then the anomalous dimension of such an operator is given by the one-loop result
\be
\gamma
= - \frac{C_2(R)- \sum_i C_2(R_i)}{2\hat{k}} \label{oneloopdim}
\ee
where $C_2$ is the quadratic Casimir defined by
\be
\sum_{\alpha}t^\alpha[R]t^\alpha[R] = C_2(R) \, \mathbf{1} \fstp
\ee
Note that for the non-Abelian group $SU(p)$, in some regularizations, there is a renormalization of $k$ to $\hat{k} = k + \rank$ due to gluon-gluon interactions. However, at leading order in $k$ this is not detectable, so it will not play a significant role in what follows. Nonetheless, we will write $\hat{k}$ in order to take account of this shift.

For example, let us return to the classically marginal interactions of an adjoint field depicted in \eqref{adjointchannels}. The singlet has the Casimir $C_2 = 0$; the adjoint has $C_2 = 2\rank$; and the antisymmetric and symmetric representations respectively have $C_2 = 4\rank - 4$ and $C_2 = 4\rank + 4$. Referring to \eqref{oneloopdim}, we find that only the symmetric channel has $\gamma < 0$. Hence, as promised, only the symmetric quartic interaction is attractive and relevant.

Now, suppose the representation $R_a$ has highest weight is $\mu$, and that we take $R$ to be the $N$th symmetric product of this representation, with highest weight $N \mu$. This is particularly natural as this representation always maximizes the quadratic Casimir $C_2(R)$ and hence minimizes \eqref{anomdimweight}, meaning this symmetric product corresponds to the lowest-dimension operator we can construct out of these $N$ particles. We can use the identity $C_2(\mu) = \left<\mu,\mu+2\rho\right>$ to obtain
\be
\gamma = - \frac{n(n-1)\left<\mu,\mu\right>}{2\hat{k}} \fstp \label{anomdimweight}
\ee
We can deduce that symmetric products of operators transforming in $R_a$ attain the unitarity bound when $\Delta = N + \gamma$ or
\be
N = \frac{2\hat{k}}{\left<\mu,\mu\right>} \fstp
\ee
Again, setting $\mu = (1)$ exactly reproduces the Abelian results we have already discussed. However, as mentioned in Chapter 5 of \cite{Turner:2017nad}, this also matches the more general result \eqref{simplenonabjpquant} for non-Abelian Jackiw-Pi vortices involving a single weight, at least up to the shift of $k \to \hat{k}$. This correspondence is natural, as $\phi_\rho^N$ necessarily transforms in the $N$th symmetric representation.

The above shows that $n_0 = 2 \hat{k}/\left<\mu,\mu\right>$ field operators combine to give an operator $\phi^{n_0}\sim\Phi$ at the unitary bound, with dimension $\Delta = 1$ -- at least if $n_0$ is an integer. But if $n_0$ is not an integer, then there is not a corresponding $\phi^{n_0}$ operator which becomes a decoupled free field. It would be interesting to check if this coincides with whether or not there is an appropriate monopole which can screen the electric charge of this operator. As noted in Appendix \ref{fusionrules}, there is such a monopole for the case of a single adjoint field where $n_0$ is indeed guaranteed to be an integer, just as there was for the Abelian case.

Arguably the more importance question is the existence of relevant operators. We can guarantee that even if $n_0$ fails to be an integer, there is an integer $r\in[n_0-1,n_0)$ such that the corresponding operator $\phi^r$ has $\Delta \in (1,2)$, at least for $k$ sufficiently large. (For integer $n_0$, the operator $\phi^{n_0-1}$ has dimension $\Delta = 2 - \left<\mu,\mu\right>/\hat{k}$, and hence gives rise to a relevant deformation for large enough $k$.)

Thus much of our understanding of the Abelian vortices more or less carries over to the non-Abelian setting -- the classical vortices arise as the classical limit of quantum operators at unitarity, and additional relevant operator appear in the spectrum to account for their size modulus. Some modifications to the argument, such as insisting we take the large $k$ limit such that $n_0$ is an integer, may be required. If we do not do so, we will still obtain generically obtain a relevant operator, but not necessarily a free particle. The story is very clear-cut for the case of a single adjoint operator, where there is an exact parallel with the Abelian case -- even down to the fact that we have only tuned a single quartic interaction.

It would be interesting to see if having multiple flavours, and hence extra moduli for the vortices, alters the analysis of operators so as to change the low-energy nature of the theory. In particular, it might prove to explore the case of multiple adjoint operators, where the additional marginal operators in the action could play a role. It would also be nice to see if there was any indication of integrability in this case.


\section*{Acknowledgements}
Thanks to Sergej Moroz, Yusuke Nishida and David Tong for useful conversations. The author is supported by a Research Fellowship in Gonville \& Caius College, Cambridge. The author also acknowledges partial support from STFC grant ST/L000385/1.


\clearpage
\appendix
\section{RG Flows for Two Abelian Anyons}\label{apprg}

In the introduction, we discussed the flow of the $\lambda$ coefficient governing the two-body interactions of Abelian anyons. This appendix serves to elaborate slightly on what this looks like.

At the stable fixed point, we have $\psi \sim r^{1/k}$, corresponding to a purely repulsive interaction, with no bound state. If one perturbs around this point, depending on the sign of $R_0$ one might introduces a very tightly bound state of small physical size and of energy $E=-E_0\ll 0$. As one follows the RG flow, we find $E_0 \to \infty$. Hence any bound state becomes increasingly tightly bound, and physically smaller, until it decouples entirely from the rest of the spectrum.

Meanwhile, at the unstable fixed point, we have $\psi \sim r^{-1/k}$. One can reach this by introducing a bound state and then tuning it to threshold, so that $R_0 \to \infty$ but $E_0 \to 0$. Alternatively, one can take $R_0 \to -\infty$. The physics is identical either way.

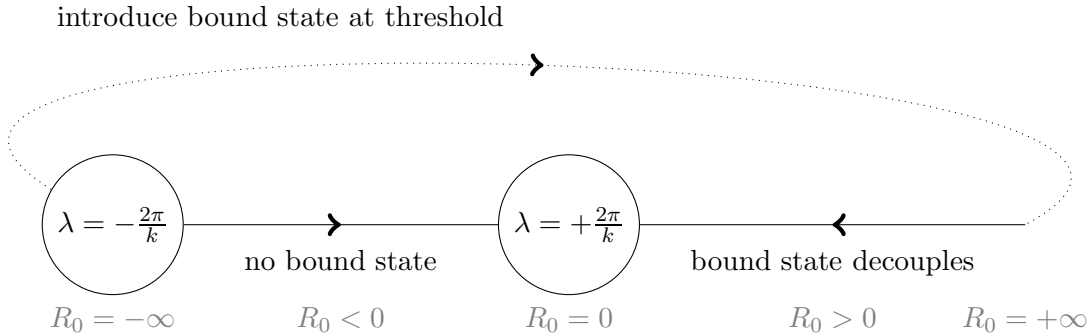
\begin{figure}[H]
	\begin{tikzpicture}
	\path[use as bounding box] (-5, -2) rectangle (11, 3);
	\node[fixedpoint] (ATT) at (-3,0) {$\lambda = -\frac{2\pi}{k}$}; 
	\node[fixedpoint] (REP) at (3,0) {$\lambda = +\frac{2\pi}{k}$}; 
	\coordinate (ATT2) at (9,0);
	\node[parameter,below=2.5em] at (ATT2) {$R_0 = +\infty$};
	\node[parameter,below=2.5em] at (REP) {$R_0 = 0$};
	\node[parameter,below=2.5em] at (ATT) {$R_0 = -\infty$};
	\draw[flow] (ATT) to node[below=0.5em]{no bound state} node[below=2.5em,parameter] {$R_0 < 0$} (REP);
	\draw[flow] (ATT2) to node[below=0.5em]{bound state decouples} node[below=2.5em,parameter] {$R_0 > 0$} (REP);
	\draw[flow,dotted] (ATT) to[bend left=150] node[above left=1em] {introduce bound state at threshold} (ATT2);
	\end{tikzpicture}
	\caption{\label{fig:anyonrg}Cartoon of the RG flow of the anyonic theory.}
\end{figure}

The final RG flow is shown in Figure \ref{fig:anyonrg}. The operator $|\phi|^4$ corresponding to deforming this boundary condition is relevant at the unstable bound state on the left with $\Delta = 4 - \frac{2}{k}$, but irrelevant at the stable bound state on the right with $\Delta = 4 + \frac{2}{k}$.

\begin{figure}[H]
	\begin{tikzpicture}
	\node[fixedpoint] (F) at (0,0) {$\lambda=0$};
	\coordinate (L) at (7,0);
	\node[parameter,below=2em] at (F) {$R_0 = 0$, $E_0 = \infty$};
	\node[parameter,below=2em] at (L) {$R_0 = \infty$, $E_0 = 0$};
	\draw[flow] (L) to node[below=0.5em]{bound state decouples} (F);
	\draw[flow,dotted] (F) to[bend left=150] node[above left=1em] {introduce bound state at threshold} (L);
	\end{tikzpicture}
	\caption{\label{fig:simplerg}Cartoon of the RG flow of the bosonic theory.}
\end{figure}
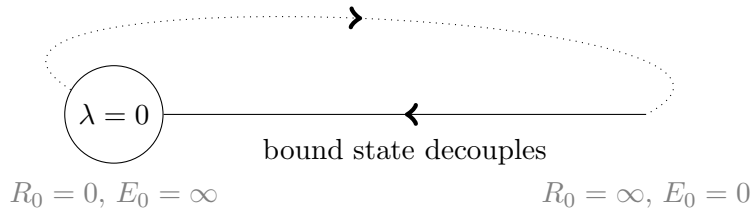

However, for the purely bosonic case, realized as $k \to \infty$, the attractive and repulsive fixed points coalesce into the non-interacting $\lambda = 0$ theory. 
In Figure \ref{fig:simplerg} we attempt to convey this information in a cartoon of the RG flow. This emphasizes that the RG flow is a form of spectral flow, in which a single bound state leaves the continuum then decouples but is replaced, Hilbert-hotel style, by another state in the continuum. Thus the fixed point, where the $|\phi|^4$ operator is marginal with $\Delta = 4$, is really a saddle point.

From an RG point of view, there is a sense in which adding a very weakly attractive interaction is a large deformation, whereas adding a very strong attractive interaction is a small one. The reason is that the former introduces a bound state of low energy, and hence is much easier to spot with low-energy experiments. However, the spectrum is discontinuous in the latter case because the infinitely tightly bound state vanishes, so from a mathematical point of view the problem behaves smoothly only when we introduce the weakly attractive interaction. We have drawn the RG flows to emphasize the physical rather than mathematical behaviour of the spectrum.

As an aside, note that the process of decoupling is very slow near the bosonic theory. One way to measure this is with the two particle scattering amplitude. The phase shift obeys
\be
\tan \delta_0 = \frac{\sin \pi/k}{\cos \pi/k - \sign R_0 (E_0/E)^{1/k}}
\label{anyonphaseshift}
\ee
in general. For the bosonic theory, this reduces to
\be
\tan \delta_0 = \frac{\pi}{\log E/E_0}  
\ee
which vanishes only logarithmically as $E_0 \to \infty$.

There is one other obvious special case to consider, namely that of $k=1$. At this point, the particles have statistical phase $\pi$: they are fermions. As emphasized in the text, fermions cannot have non-trivial contact interactions, and the only possible interpretation available for the second fixed point is that there are decoupled two-particle bound states as well as the usual fermionic spectrum. This suggests that the RG flow for the two-particle system becomes somewhat trivial at $k=1$. The stable fixed point is simply a free fermion. The unstable fixed point is a free fermion and a free boson, tuned to have energy 0. The RG flow between them simply consists of the energy of bosonic particles going to $\pm\infty$ so that the boson decouples.

This gives a natural understanding of the RG flow for all $k\in[1,\infty]$ -- things are essentially the same for all values of $k$. Of course, things can be more interesting with more particles, and that is what is discussed in section \ref{smallksec}.

\section{A Note on Fusion Rules}\label{fusionrules}
One important feature of Chern-Simons theories is their \textit{fusion rules} which dictate the possible representations of the gauge group in which local operators may exist (the distinct species of anyons), and how these combine in the OPE. For $SU(\rank)_k$, for instance, all so-called \textit{integrable} representations may be labelled by a Young diagram with at most $k$ columns. Since we are interested precisely in operators with symmetrizations of order $k$ operators, we should try to understand how the fusion rules interact our story.

From the point of view of pure Chern-Simons theory, the simplest way to understand this is to look at \eqref{nonabgauss}. This equation indicates that a monopole operator carrying magnetic charge $Q^\alpha = \frac{1}{2\pi}\int f^\alpha$ is imbued with an electric charge $kQ^\alpha$. Including these operators in the path integral can screen the charges due to matter fields; and taking careful account of the signs of the path integral one determines that summing over these magnetic charges actually projects out certain states from the physical Hilbert space. (Equivalently, in the presence of a Maxwell term with gauge coupling $e^2$, these states are given  $O(e^2)$ masses.) One is left with a subset of the local operators one might na\"ively have expected. At the stable fixed point, the way this works is very natural, and is discussed in e.g. \cite{Turner:2017rki}. However, things appear different at the unstable fixed point.

We will not discuss this in detail; but we will note that it is perfectly possible that states hitting unitarity are not projected out by this process. This is especially plausible if there are no other states of the same dimension and particle number for a state to mix with, as is the case for the first operator to hit unitarity in our discussions. However, the UV-divergent nature of the state may require careful consideration of gauge field dynamics.

As an example we have already seen, something very neat happens in the $U(1)_k$ case: the relevant $2k$-particle state can be screened by monopole flux so that the many-body quantum state transforms as a gauge singlet. Note that the fusion rules for $U(1)_k$ Chern-Simons theory are very simple. Representations with $U(1)$ charge $N$ have spin $N^2/2k$. For $k$ even, there are simply $k$ distinct allowed representations, arising from $0\le N <k$; a state with $N=k$ would be bosonic, not anyonic. For $k$ odd, we must take $0 \le N < 2k$, as the $N=k$ state has spin $k/2$ and so is fermionic. Hence for generic $k$, only $N=2kq$ states are necessarily bosonic in character; precisely the operators which hit unitarity.

For the case of a single $SU(\rank)_k$ adjoint, which seems particularly amenable to study, we know that the $\hat{k}$th symmetric product is the first to hit unitarity. A careful examination of the fusion rules reveals that screening by flux allows this to transform as a gauge singlet for all values of $\rank,k$. This may well be related to the particularly nice properties of the classical equations for these theories. Indeed, the classical uniton solutions exist when space is compactified, so that Gauss's law enforces that physical states are gauge singlets.

It seems reasonable to conjecture a deeper connection between fusion rules and unitarity bounds; perhaps states \textit{exactly} saturating unitarity bounds should transform as gauge singlets so that they can be realized as free fields. (As circumstantial evidence: the quadratic Casimir $C_2(R)$ appears in formulae for both the spin and dimension of operators.)


\clearpage
\bibliography{biblio}


\end{document}